\documentclass[11pt]{article}  
\usepackage{graphicx}% Include figure files 
\newcommand{\be}{\begin{equation}}
\newcommand{\ee}{\end{equation}}
\newcommand{\bea}{\begin{eqnarray}}
\newcommand{\eea}{\end{eqnarray}}

\begin{document}
\vspace{0.5in}
\vspace{.5in} 
\begin{center} 
{\LARGE{\bf Novel Deformation Function Creating or Destroying any Number
of Even Kink Solutions}}
\end{center} 
\vspace{0.2in} 

\begin{center}
{{\bf Avinash Khare}} \\
{Physics Department, Savitribai Phule Pune University \\
 Pune 411007, India}
\end{center}

\begin{center}
{{\bf Avadh Saxena}} \\ 
{Theoretical Division and Center for Nonlinear Studies, 
Los Alamos National Laboratory, Los Alamos, New Mexico 87545, USA}
\end{center}

\vspace{0.2in}

\noindent{\bf {Abstract:}}
We present a one-parameter family of deformation functions $f(\phi)$ which have 
novel properties. Firstly, the deformation function is its own inverse. We show that 
a class of potentials remains invariant under this deformation. 
Further, when applied to a certain class of kink bearing potentials, one
obtains potentials which are not bounded from below. Besides, we show that
there is a wide class of potentials having power law kink tails which are 
connected by this deformation but the corresponding kink tails of the two
potentials have different asymptotic behavior. Finally, we show that when
this deformation function is applied to an appropriate one-parameter family of 
potentials having two kink solutions, it creates new potentials with an arbitrary 
even number ($2n$) of kink solutions. Conversely, by starting from this potential 
with $2n$ kink solutions the deformation function annihilates $2n-2$ kinks and we 
get back the potential which we started with having only two kink solutions.

\section{Introduction} 
%Recently a few analytic kink solutions have been obtained for the $\phi^8$ field theory \cite{ganiPRD}.  It opens up the possibility that certain exact kink solutions may also 
%exist under specific constraints for the $\phi^{10}$, $\phi^{12}$ and even higher order filed theories \cite{KCS, Chapter}.  ... 
During the last few years there has been a resurgence in studying kink 
solutions with power-law tails in several higher (than 6th) order field theory 
models including $\phi^8$, $\phi^{10}$, and $\phi^{12}$ \cite{CSK, Chapter} 
and even $\phi^{14}$, $\phi^{16}$, and $\phi^{18}$ \cite{KS20}. 
The study of higher order field theories, their attendant kink excitations as 
well as the associated kink interactions and scattering are important in a 
variety of physical contexts ranging from successive phase transitions 
\cite{CSK, Chapter, Gufan82, Gufan78} to isostructural phase transitions 
\cite{Pavlov99} to models involving long-range interaction between massless 
mesons \cite{Lohe79}, as well as from protein crystallization 
\cite{Boulbitch97} to successive phase transitions presumably driving the late 
time expansion of the Universe \cite{Greenwood09}. Thus, it is of interest to
study kink solutions in additional higher order field theories and study
the various properties of these solutions including the nature of the 
kink-kink and the kink-antikink forces and scattering in these theories. 

In this context we mention that almost two decades ago, in an interesting paper, 
Bazeia, Losano and Malbonisson \cite{blm} introduced the idea of deformation 
function which takes one from a kink bearing potential to another kink bearing 
(or in some cases pulse bearing) potential. In a subsequent series of papers, 
these and other authors \cite{ablm, bl, bllg, bgg} studied a variety of deformation 
functions and also discussed possible applications of some of these deformation 
functions.

We mention in particular the deformation function $f(\phi) = (1-\phi^2)^{1/2}$ 
introduced in \cite{bl} where the authors briefly discussed some of its properties. 
The purpose of this paper is to discuss a one-parameter family of deformation functions 
\be\label{1}
f(\phi) = (1-\phi^{2n})^{1/2n}\,,~~n = 1, 2, 3,... \,,
\ee
where $n$ is any positive integer. In particular we point out its several 
remarkable properties. 

The plan of the paper is the following. In Sec. II we 
set up the formalism for the kink solutions in a neutral scalar field theory 
in $1+1$ dimensions and then spell out the basic steps of the deformation 
prescription. We then discuss some of the interesting properties of the 
deformation function (\ref{1}).
In particular, if under the deformation (\ref{1}), a scalar potential 
$V_0(\phi)$ goes to another potential $V_1(\phi)$, then the same deformation
function acting on $V_1(\phi)$ gives back the original potential $V_0(\phi)$ 
i.e. $V_0$ and $V_1$ are deformation-dual potentials. In other words, the
deformation (\ref{1}) is inverse of itself. We then exhibit a wide class of
potentials $V_0(\phi)$ which are invariant under this deformation; we call 
such potentials as self-deformed potentials. Further, we exhibit a wide class 
of kink bearing potentials $V_0(\phi)$ which under the deformation (\ref{1})
go to potentials $V_1(\phi)$ which are not bounded from below. In Sec. III
we then discuss a wide class of deformation-dual potentials both of which
admit 2 kink solutions and show that in case at least one of the kink tails 
of $V_0$ and $V_1$ is a power law tail, then the power law tail behavior
of $V_0$ and the corresponding tail of $V_1$ are different. This is unlike
most of the examples discussed so far in the literature. We illustrate this
feature by discussing one example in detail. In Sec. IV we exhibit
perhaps the most remarkable property of the deformation (\ref{1}). In 
particular, we show that by appropriately choosing the potential 
$V_0(\phi)$ with two kink solutions and by applying the deformation (\ref{1}), 
one can obtain the potential $V_1$ with as many even number $2m$ of kink
solutions as desired. We discuss one example in detail to illustrate this aspect. 
Needless to say that, if instead we apply the deformation (\ref{1}) on $V_1$
having $2m$ kink solutions, we obtain the potential $V_0$ with two kink
solutions. In other words, by an appropriate choice of the potential, one can
create or destroy as many even number of kink solutions as desired. Finally,
in Sec. V we summarize the main results obtained in this paper and point out 
a few open questions.

\section{Formalism and Important Properties of the Deformation Function}

Consider a relativistic neutral scalar field theory in $1+1$ dimensions with
the Lagrangian density being \cite{raj}
\be\label{2.1}
{\rm{\cal{L}}} = \frac{1}{2} \left(\frac{d\phi}{dt}\right)^2 
- \frac{1}{2} \left(\frac{d\phi}{dx}\right)^2 -V(\phi)\,,
\ee
which leads to the equation of motion
\be\label{2.2}
\frac{d^2\phi}{dt^2} - \frac{d^2\phi}{dx^2} = \frac{dV}{d\phi}\,.
\ee
We assume that the potential $V(\phi)$ is smooth and nonnegative and 
attains its global minimum value of $V = 0$ for two or more values of 
$\phi$ which are the global minima of the theory. Thus one has static kink
and antikink solutions interpolating between the two adjoining global minima
as $x$ goes from $-\infty$ to +$\infty$. While the field equation for a static 
kink is a second order ODE, it can be reduced to a first order ODE using the
so called Bogomolnyi technique \cite{Bogom}. The first order ODE is given by
\be\label{2.3}
\frac{d\phi}{dx} = \pm \sqrt{2 V(\phi)}\,.
\ee

Let us now point out a few basic properties of the deformation formalism 
\cite{blm, bgg}. In particular, if we start with a potential $V_0(\phi)$ and 
apply a deformation function $f(\phi)$, then the new potential $V_1(\phi)$ 
is related to  the original potential $V_0(\phi)$ by
\be\label{2.4}
V_1(\phi) = \frac{V_0(\phi)}{\left[\frac{df(\phi)}{d\phi}\right]^2}\,.
\ee
Further, if the explicit kink solutions corresponding to the two potentials 
$V_0(\phi)$ and $V_1(\phi)$ are $\phi^{(0)}_{K}(x)$ and $\phi^{(1)}_{K}(x)$ 
respectively, then they are related by
\be\label{2.5}
\phi^{(1)}_{K}(x) = f^{-1}[\phi^{(0)}_{K}(x)]\,.
\ee
On the other hand, if instead of the explicit kink solution, only the implicit
kink solutions are known and if corresponding to the potentials $V_0(\phi)$ and
$V_1(\phi)$ they are given by $x^{(0)}_K(\phi)$ and $x^{(1)}_{K}(\phi)$ 
respectively, then
\be\label{2.6}
x^{(1)}_K(\phi) = x^{(0)}_K(f[\phi])\,.
\ee

Let us now discuss the various properties of the deformation function  
given by Eq. (\ref{1}).

\begin{enumerate}

\item Notice that the inverse of this deformation is equal to itself, i.e.
\be\label{2.7}	
f^{-1}(\phi) = f(\phi)\,.
\ee
Stated differently, if we start with a potential $V_0(\phi)$ and after applying
the deformation function $f(\phi)$ as given by Eq. (\ref{1}) if we get the 
potential $V_1(\phi)$, then we know that $V_0$ and $V_1$ are related to each 
other as given by Eq. (\ref{2.4}). Similarly, if we apply the same deformation 
function to $V_1(\phi)$, then we get back the potential $V_0(\phi)$, i.e.
\be\label{2.8}		
V_0(\phi) = \frac{V_1(\phi)}{\left[\frac{df(\phi)}{d\phi}\right]^2}\,, 
\ee
where $f(\phi)$ is given by Eq. (\ref{1}). In that case, we call $V_0$ and
$V_1$ as deformation-dual or deformation pair. 
As a result, the relation between the explicit kink solutions of the potentials
$V_0(\phi)$ and $V_1(\phi)$ given by Eq. (\ref{2.5}) can also be expressed as
\be\label{2.9}
\phi^{(1)}_{K}(x) = f[\phi^{(0)}_{K}(x)]\,.
\ee

\item For every integer $n$, a wide class of potentials are invariant under 
the deformation (\ref{1}). We call such potentials as self-dual potentials. 
Using Eq. (\ref{1}) it is easy to check that for an arbitrary integer $n$, the potential
\be\label{2.10}
V_0(\phi) = \frac{1}{2} \phi^{4nm+2} [1-\phi^{2n}]^{2m+2} = V_1(\phi)\,,~~m
= 0, 1, 2,... \,.
\ee
We term these wide classes of potentials as self-deformed potentials.
As an illustration, in the case of $n = 1$, the self-deformed one parameter
family of potentials is
\be\label{2.11}
V_0(\phi) = \frac{1}{2} \phi^{4m+2} [1-\phi^2]^{2m+2} = V_1(\phi)\,,~~m =
0, 1, 2,...\,.
\ee

\item For every integer $n$, there is a two-parameter family of kink bearing 
potentials $V_0$ for which when we apply the kink deformation function 
(\ref{1}) to them, the resulting deformed-dual potential
$V_1$ is not bounded from below. In particular, for an arbitrary integer $n$, 
consider the two-parameter family of potentials
\be\label{2.12}
V_0(\phi) = \frac{1}{2} \phi^{4 n m +2(2n-1)(n-1)} (1-\phi^{2n})^{2p+2}\,,
~~~m, p = 0, 1, 2,... \,,
\ee 
and if we apply the deformation function (\ref{1}) to this $V_0$, then we find
that the corresponding potential $V_1$ is given by 
\be\label{2.13}
V_1(\phi) = \frac{1}{2} \phi^{4np+2} (1-\phi^{2n})^{2m+2n-1}\,. 
\ee 
Notice that $V_1(\phi)$ is not bounded from below. As an illustration, consider
the special case when $n = m = 1, p =0$, in which case the deformed-dual 
potentials are
\be\label{2.14}		
V_0(\phi) = \frac{1}{2} \phi^4 (1-\phi^2)^2\,,~~~V_1(\phi) 
= \frac{1}{2} \phi^2 (1-\phi^2)^3\,.
\ee
It is worth pointing out that the potential $V_0$ in Eq. (15) has recently 
received 
a lot of attention \cite{Lohe79, ks19, man19, Radomskiy17, cdd, cddg} in the 
context of kink solutions with a power law tail around $\phi = 0$ while an 
exponential tail around $\phi = \pm 1$. As is clear from Eq. (\ref{2.14}), the 
corresponding dual potential $V_1(\phi)$ is not bounded from below.

\end{enumerate}

\section{Comparison of Kink Tails of the Deformed-Dual Potentials} 

We now consider for any arbitrary integer $n$, a two-parameter family of 
deformed-dual potentials given by 
\be\label{3.1}
V_0(\phi) = \frac{1}{2} \phi^{4np+2} (1-\phi^{2n})^{2m+2}\,,
~~p, m = 0, 1, 2,... \,,
\ee
and
\be\label{3.2}
V_1(\phi) = \frac{1}{2} \phi^{4nm+2} (1-\phi^{2n})^{2p+2}\,.
\ee
Note that here $p \ne m$ since in that case the two potentials $V_0, V_1$ 
are one and the same, i.e. it is the self-deformed potential. We note that 
for any set of integers $p, m, n$, both the deformed-dual potentials (\ref{3.1}) and 
(\ref{3.2}) have two kinks and two antikink solutions. 
We now show that (for $p \ne m$) at least one of the kink (and hence the 
antikink) tails of 
$V_0$ and the corresponding kink tails of $V_1$ have different asymptotic
behavior. In this context it is worth reminding that for most of the examples
discussed in the literature, the kink tails of $V_0$ and the corresponding
kink tails of $V_1$ are the same up to an overall factor.

In this context let us note the well known result that as $x$ goes 
from $-\infty$ to +$\infty$, if a kink solution for a potential $V(\phi)$ 
goes from $0$ to $b$ and if $V(\phi)$ behaves as $\phi^q$
around $\phi = 0$, then the kink tail around $\phi = 0$ has the asymptotic
behavior 
\be\label{3.3}
\lim_{x \rightarrow -\infty} \phi_{K}(x) = x^{-(q-2)/2}\,,
\ee
where $q$ is a real number. Thus if  $q = 2$ then the kink tail is exponential 
while for $q > 2$ the kink has a power law tail. 
Let us now use this result and look at the behavior of the kink
solution around $\phi = 0$ and $\phi = 1$ for the dual potentials given above.
Using Eq. (\ref{3.3}), it is straightforward to show that for the potential
$V_0(\phi)$ as given by Eq. (\ref{3.1}), for the kink solution between $0$ and 
$1$, the behavior of the kink tail is given by
\be\label{3.4}
\lim_{x \rightarrow -\infty} \phi^{(0)}_{K}(x) = \frac{A_0}{x^{4p}}\,,~~~
\lim_{x \rightarrow \infty} \phi^{(0)}_{K}(x) = 1- \frac{A_1}{x^{m}}\,.
\ee
On the other hand, for the potential $V_1(\phi)$ as
given by Eq. (\ref{3.2}), for the kink solution between $0$ and $1$, 
the behavior of the kink tail is given by
\be\label{3.5}
\lim_{x \rightarrow -\infty} \phi^{(0)}_{K}(x) = \frac{B_0}{x^{4m}}\,,~~~
\lim_{x \rightarrow \infty} \phi^{(0)}_{K}(x) = 1- \frac{B_1}{x^{p}}\,.
\ee
Here $A_0, A_1, B_0, B_1$ are constants. Since under the deformation (\ref{1}),
while $\phi = 0$ and $\phi = 1$ of $V_0(\phi)$ get mapped to $\phi = 1$ 
and $\phi = 0$ respectively of $V_1(\phi)$, one would expect that the behavior 
of the kink tail around $\phi = 0 (1)$ for the potential $V_0(\phi)$ should 
be the same as that of the kink tail around $\phi = 1 (0)$ 
for the deformation-dual potential $V_1(\phi)$. However it is clear from Eqs. (\ref{3.4}) 
and (\ref{3.5}) that this is not so and the asymptotic behavior of the kink tails in the  
two dual potentials (\ref{3.1}) and (\ref{3.2}) is very different  unless $p = m$ but in 
that case the two potentials are self-deformed (i.e. $V_1(\phi) = V_0(\phi)$). 
It is worth pointing out that in case, say $p = 0$ but $m \ne 0$, then the kink 
tail of $V_0(\phi)$ around $\phi = 0$ and the kink tail of $V_1(\phi)$ around 
$\phi = 1$ are both exponential while the kink tail of $V_0(\phi)$ around 
$\phi = 1$ and the kink tail of $V_1(\phi)$ around $\phi = 0$ are both power
law tails having different asymptotic behavior. On the other hand when both
$p, m > 0$ but $p \ne m$, then both the kink tails of $V_0(\phi)$ are different
from the corresponding kink tails of $V_1(\phi)$, 

For simplicity we now explicitly discuss the case of $n = m = 1, p = 0$ and 
demonstrate that indeed the appropriate kink tails of the two deformed-dual
potentials are different. In case $n = m = 1$, $p = 0$,
$V_0(\phi)$ and $V_1(\phi)$ are given by
\be\label{3.6}
V_0(\phi) = \frac{1}{2} \phi^{2} (1-\phi^{2n})^{4}\,,
\ee
\be\label{3.7}
V_1(\phi) = \frac{1}{2} \phi^{6} (1-\phi^{2n})^{2}\,.
\ee
These two potentials are depicted in Fig. 1 with three degenerate minima each. 
The kink solutions in both the cases can be implicitly obtained. 

\begin{figure}[h] 
\includegraphics[width=6.0 in]{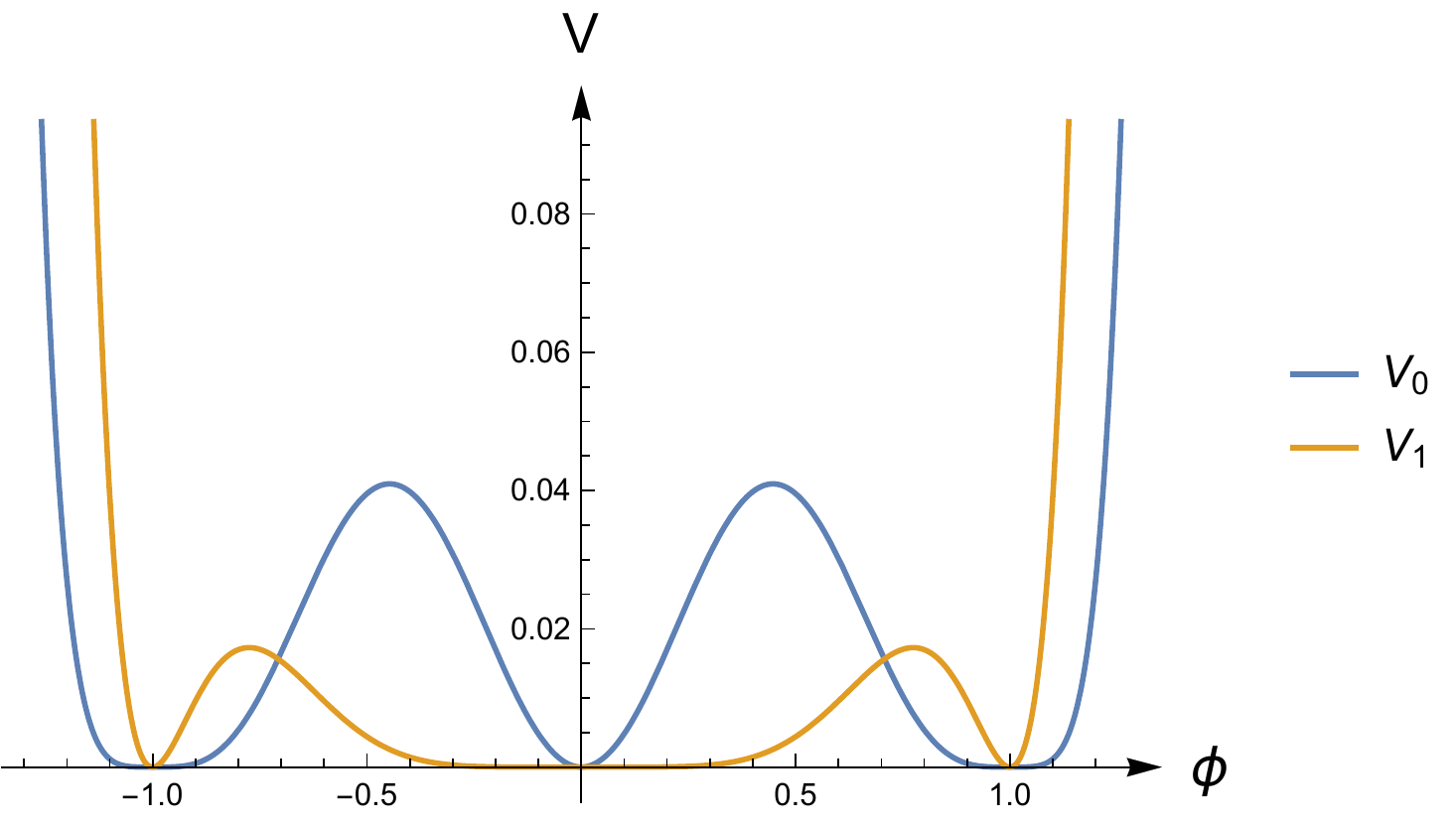}
\caption{Potential $V_0(\phi)$ in Eq. (21) and its dual potential $V_1(\phi)$ in Eq. (22) for $n=1$.  }
%\label{KS1}
\end{figure} 

For example, in order to obtain the kink solution of $V_0(\phi)$ as given by
Eq. (\ref{3.6}) from $0$ to $1$, we need to solve the self-dual equation
\be\label{3.8}
\frac{d\phi}{dx} = \phi (1-\phi^2)^2\,.
\ee
This equation is easily integrated using partial fractions and we obtain
\be\label{3.9}
x^{(0)} = \frac{1}{2(1-\phi^2)} + \frac{1}{2} \ln\left[\frac{\phi^2}{(1-\phi^2)}\right]\,.
\ee
It then immediately follows that for the potential $V_0(\phi)$ as given by 
Eq. (\ref{3.6}), the asymptotic behavior of the kink solution from $0$ to
$1$ is given by 
\be\label{3.10}
\lim_{x \rightarrow -\infty} \phi^{(0)}_{K}(x) = e^{(x-1/2)}\,,~~
\lim_{x \rightarrow \infty} \phi^{(0)}_{K}(x) = 1- \frac{1}{4 x}\,.
\ee
The kink solution given by Eq. (24) is shown in Fig. 2 with the exponential and 
power law tails clearly visible at the two ends. 

\begin{figure}[h] 
\includegraphics[width=6.0 in]{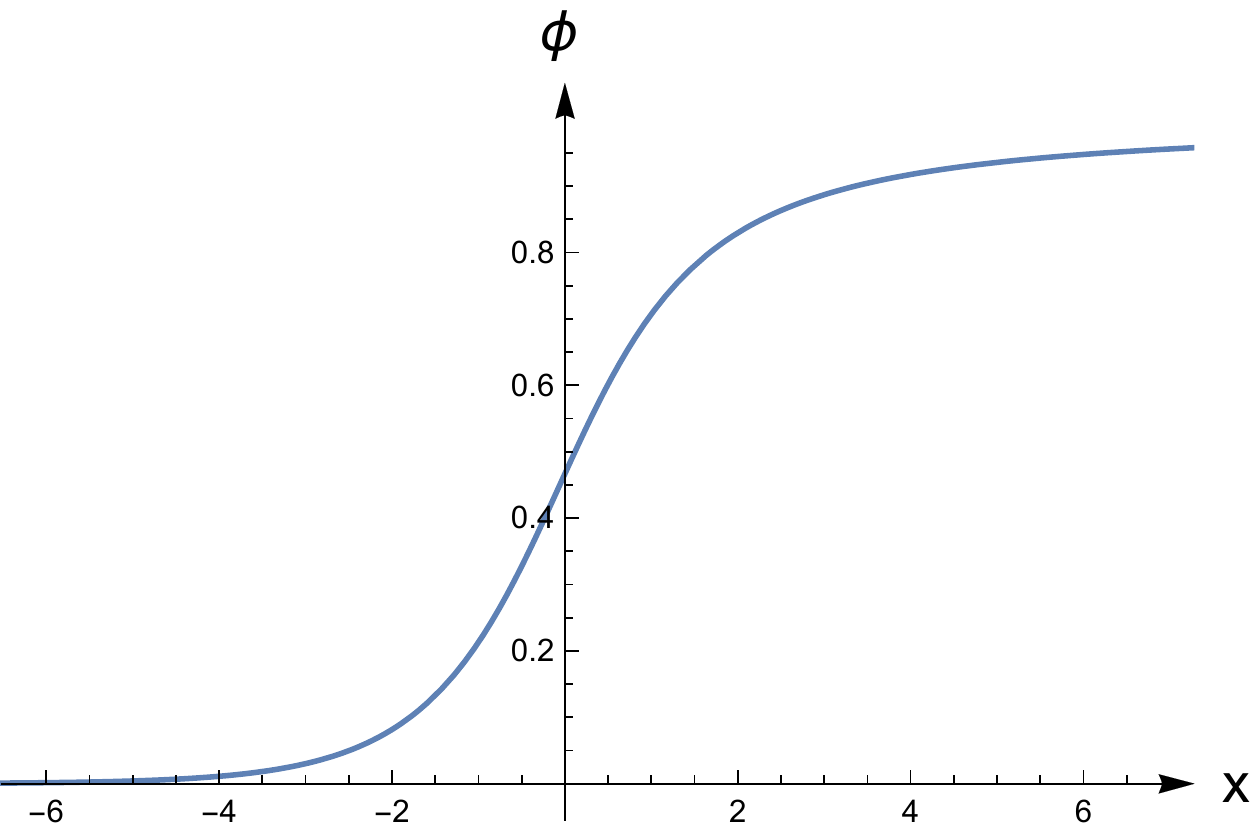}
\caption{Kink solution obtained by inverting implicit Eq. (24).  }
%\label{KS1}
\end{figure} 

In order to obtain the kink solution of $V_1(\phi)$ as given by Eq. (\ref{3.7})
from $0$ to $1$, we need to solve the self-dual equation
\be\label{3.11}
\frac{d\phi}{dx} = \phi^3 (1-\phi^2)\,.
\ee
This equation is easily integrated using partial fractions and we obtain
\be\label{3.12}
x^{(1)} = -\frac{1}{2\phi^2} + \frac{1}{2} \ln\left[\frac{\phi^2}{(1-\phi^2)}\right]\,.
\ee
It then immediately follows that for the potential $V_1(\phi)$ as given by 
Eq. (\ref{3.7}), the asymptotic behavior of the kink solution from $0$ to
$1$ is given by 
\be\label{3.13}
\lim_{x \rightarrow -\infty} \phi^{(1)}_{K}(x) = \frac{1}{\sqrt{-2x}}\,,~~~
\lim_{x \rightarrow \infty} \phi^{(1)}_{K}(x) = 1- \frac{1}{2} e^{-2(x+1/2)}\,,
\ee
so that while the kink tail of $V_0$ goes like $x^{-1}$ around $\phi = 1$, the
kink tail of $V_1$ around $\phi = 0$ goes like $(-x)^{1/2}$.  The kink solution 
given by Eq. (27) is shown in Fig. 3 with the power law and exponential tails 
clearly visible at the two ends. Needless to say that alternatively we can also 
calculate the kink tails of the potential (\ref{3.7}) by using Eqs. (\ref{2.6}) and 
(\ref{3.9}) and we would get the same kink tails as given by Eq. (\ref{3.13}) up 
to an overall factor. 
\begin{figure}[h] 
\includegraphics[width=6.0 in]{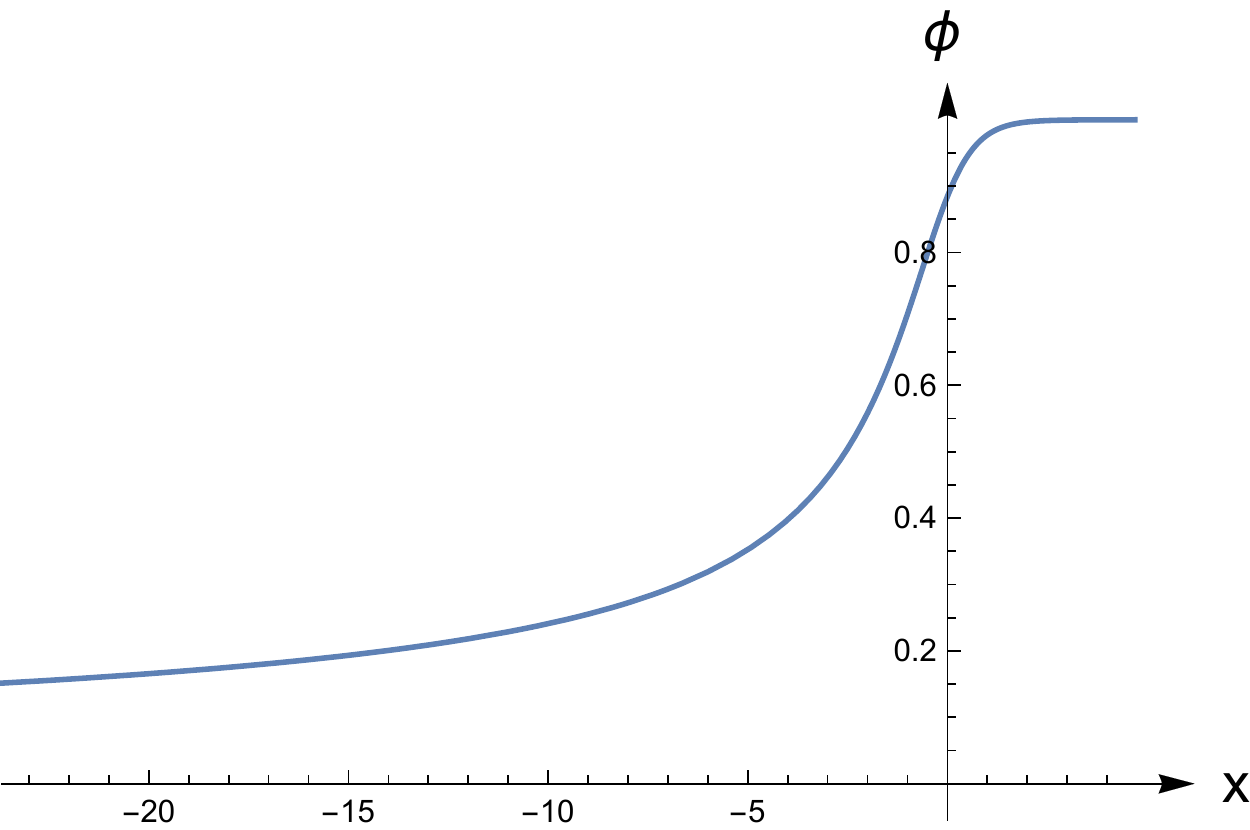}
\caption{Kink solution obtained by inverting implicit Eq. (27).   }
%\label{KS2}
\end{figure}   

\section{Kink Creating and Kink Annihilating Deformation}

Finally, we discuss perhaps the most interesting property of the deformation 
function (\ref{1}). In particular, by a judicious choice of the potential $V_0(\phi)$ 
having two kink solutions, this deformation can create a new potential $V_1(\phi)$ 
having as many  even number of kink solutions as desired. We give below three 
wide classes of $V_0$ and $V_1$ satisfying this property.

Let us start from the potential
\be\label{4.1}
V_0(\phi) = \frac{1}{2} \phi^2 (1-\phi^{2n})^2 (b_{1}^{2n}+ \phi^{2n})^2 
(b_{2}^{2n}+ \phi^{2n})^2 (b_{3}^{{2n}}+ \phi^{2n})^2 ...
(b_{m}^{2n} +\phi^{2n})^2\,,
\ee 
which has two kink and two antikink solutions. On applying the deformation
function (\ref{1}) on this $V_0(\phi)$ we obtain the deformed potential
\be\label{4.2}
V_1(\phi) = \frac{1}{2} \phi^2 (1-\phi^{2n})^2 (b_{1}^{2n}+1 - \phi^{2n})^2 
(b_{2}^{2n}+1 - \phi^{2n})^2...(b_{m}^{2n}+1-\phi^{2n})^2\,,
\ee 
which clearly has $2m+2$ kink solutions and $2m+2$ antikink solutions.
Thus in this case, for $V_0(\phi)$ as given by Eq. (\ref{4.1}), the deformation 
function (\ref{1}) acts like a kink (and antikink) creating operator, creating extra 
$2m$ kink solutions. From here it immediately follows that, as far as the potential 
$V_1(\phi)$ as given by Eq. (\ref{4.2}) is concerned, the deformation function 
(\ref{1}) acts like a kink (and antikink) annihilating operator destroying $2m$ 
kinks out of its $2m+2$ kinks.

One can generalize the potential $V_0$ given by Eq. (\ref{4.1}) and
choose more general potentials of the form
\be\label{4.3}
V_0(\phi) = \frac{1}{2} \phi^2 (1-\phi^{2n})^2 (b_{1}^{2n}+ \phi^{2n})^{2p_1} 
(b_{2}^{2n}+ \phi^{2n})^{2p_2}...(b_{m}^{2n} +\phi^{2n})^{2p_m}\,,
\ee 
where $p_1, p_2, ..., p_m$ are arbitrary positive integers. This $V_0$ 
obviously has has two kink and two antikink solutions. On applying the deformation
function (\ref{1}) on this $V_0(\phi)$ we obtain the deformed potential
\be\label{4.4}
V_1(\phi) = \frac{1}{2} \phi^2 (1-\phi^{2n})^2 
(b_{1}^{2n}+1 - \phi^{2n})^{2p_1} 
(b_{2}^{2n}+1 - \phi^{2n})^{2p_2}...(b_{m}^{2n}+1-\phi^{2n})^{2p_m}\,,
\ee 
which clearly has $2m+2$ kinks and $2m+2$ antikink solutions. 

We can further generalize the potential $V_0(\phi)$ of Eq. (31) and choose 
more general potentials of the form 
\be\label{4.5}
V_0(\phi) = \frac{1}{2} \phi^{4np+2} (1-\phi^{2n})^{2k+2}
(b_{1}^{2n}+ \phi^{2n})^{2p_1} 
(b_{2}^{2n}+ \phi^{2n})^{2p_2}...(b_{m}^{2n} +\phi^{2n})^{2p_m}\,,
\ee 
where $p,k = 0, 1, 2,... $ while $p_1, p_2,...,p_m$ are arbitrary positive integers. 
Clearly this $V_0$ has two kinks and two antikink solutions.
On applying the deformation function (\ref{1}) on this $V_0(\phi)$, we obtain
the deformed potential
\be\label{4.5a}
V_1(\phi) = \frac{1}{2} \phi^{4nk+2} (1-\phi^{2n})^{2p+2}
(b_{1}^{2n}+1 - \phi^{2n})^{2p_1} (b_{2}^{2n}+1 - \phi^{2n})^{2p_2}...
(b_{m}^{2n}+1-\phi^{2n})^{2p_m}\,,
\ee 
which clearly has $2m+2$ kink solutions.

Remarkably, there are several potentials $V_0$ and $V_1$ in which kink 
solutions are explicitly or implicitly known and they fall in the above
category. As an illustration we discuss two such examples, in one of which 
while $V_0$ has two kink solutions which are explicitly known, the corresponding
$V_1$ has four kink solutions which are all explicitly known as well. In another case, 
while $V_0$ has two kink solutions which are implicitly known, the corresponding 
$V_1$ has six kink solutions all of which are implicitly known as well. 

\vskip 0.1truein 
\noindent{\bf Example I:}

Few years ago, we along with Christov \cite{Chapter} had explicitly obtained
two kink and two antikink solutions for the one-parameter family of potentials 
\be\label{4.6}
V_0(\phi) = \frac{1}{2} \phi^2 (1-\phi^{4n})^2\,,~~n = 1, 2, 3,... \,.
\ee
On the other hand, very recently, two of us \cite{ks21} have obtained explicit 
four kink and four antikink solutions of the model 
\be\label{4.7}
V_1(\phi) = \frac{1}{2} \phi^2 (1-\phi^{2n})^2 (2-\phi^{2n})^2\,,~~
n = 1, 2, 3,... \,.
\ee
Note that these $V_0$ and $V_1$ are in fact special cases of the potentials 
given in Eqs. (\ref{4.1}) and (\ref{4.2}), respectively, in case $b_1 = 1$ 
while $b_2 = ...= b_m =0$ and hence $V_0(\phi)$ and $V_1(\phi)$ as given by 
Eqs. (\ref{4.6}) and (\ref{4.7}) are deformation-dual.  The potentials given by Eqs. 
(35) and (36) for $n=1$ are depicted in Fig. 4 with three and five degenerate 
minima, respectively. 
\begin{figure}[h] 
\includegraphics[width=6.0 in]{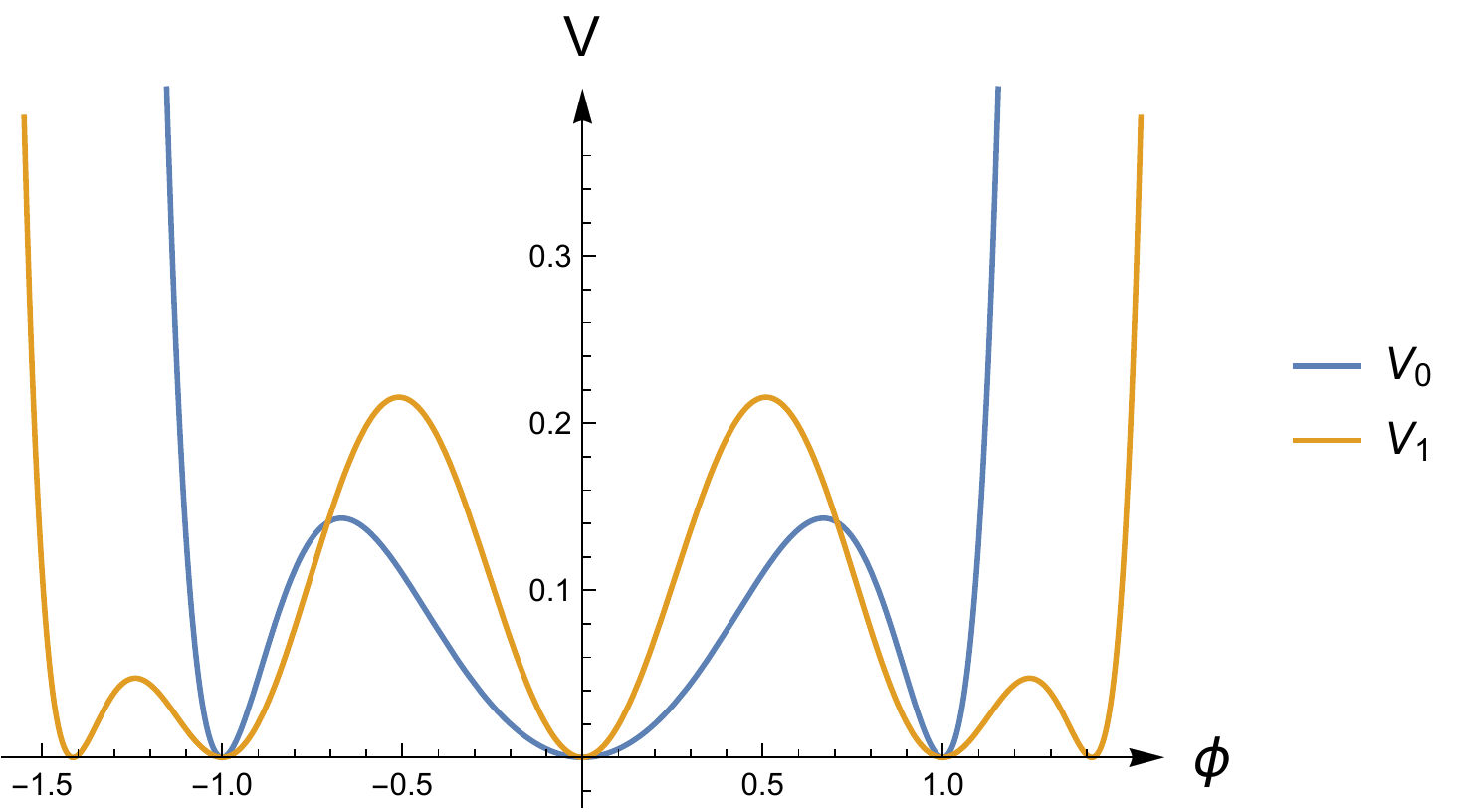}
\caption{Potential $V_0(\phi)$ in Eq. (35) and its dual potential $V_1(\phi)$ in Eq. (36) for $n=1$.  }
%\label{KS2}
\end{figure}  

As has been shown in \cite{CSK}, for the potential (\ref{4.6}) with arbitrary
$n$, the kink from $0$ to $1$, the corresponding mirror kink from $-1$
to $0$ and the corresponding two antikink solutions are given by
\be\label{4.8}
\phi_{0K,0aK}(x) = \pm \left[\frac{1\pm \tanh(2n x)}{2}\right]^{1/4n} \,. 
\ee
For $n=1$ this kink solution is depicted in Fig. 5. Having obtained all the kink and 
the antikink solutions of the potential (\ref{4.6}) from $0$ to $1$, the corresponding 
kink, mirror kink and the two anti-kink solutions of the dual potential (\ref{4.7})  
are then immediately obtained from the deformation formalism. In particular 
using Eq. (\ref{2.9}) we obtain 
\be\label{4.9}
\phi_{1K,1aK}(x) = \pm [1-\phi^2_{0K,0aK}(x)]^{1/4n} = 
\pm \left[1-\sqrt{\frac{1\pm \tanh(2n x)}{2}}\right]^{1/4n}\,.
\ee
For $n=1$ his kink solution is depicted in Fig. 6. It is worth pointing out that these 
are precisely the explicit kink solutions that we have recently obtained \cite{ks21} 
for the potential (\ref{4.7}). Thus, one has found a novel relationship between the 
two seemingly very different potentials, one admitting two while the other admitting 
four kink solutions.
\begin{figure}[h] 
\includegraphics[width=6.0 in]{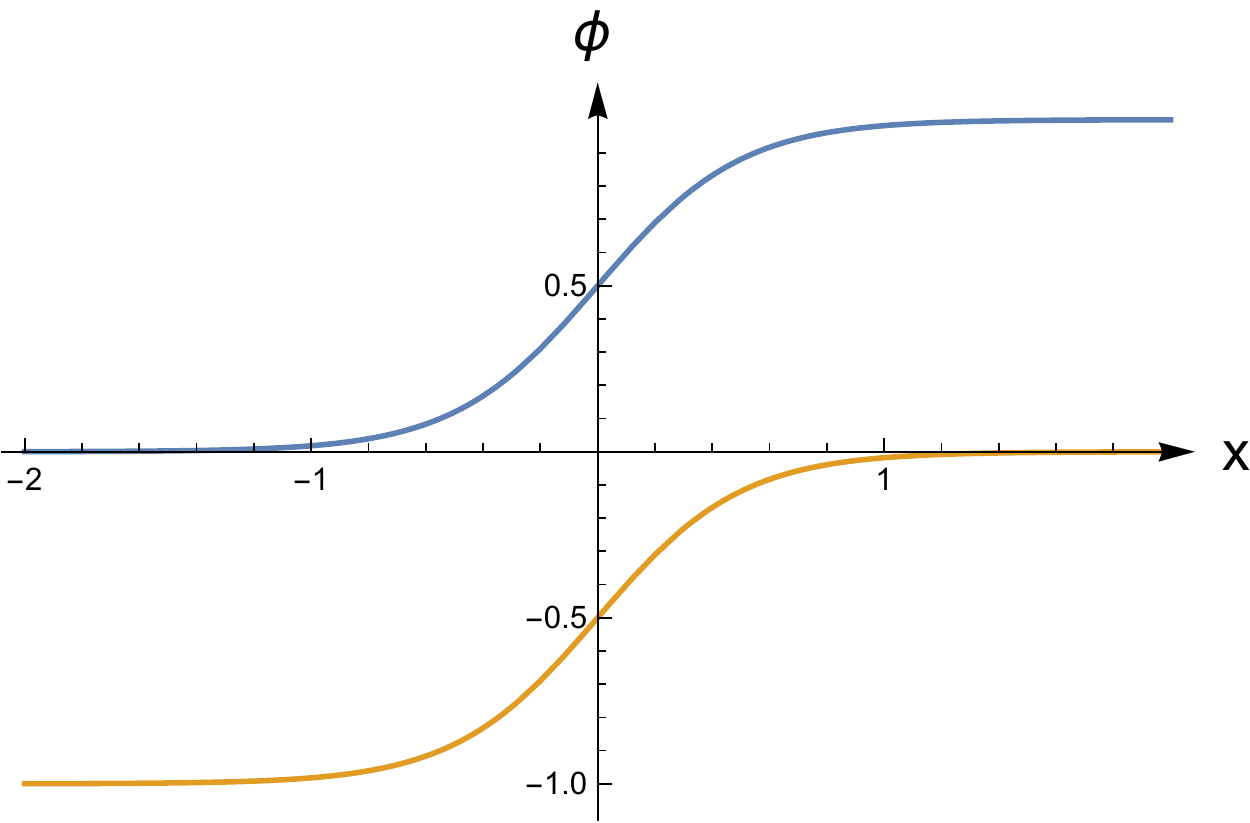}
\caption{Explicit kink solution given by Eq. (37) for $n=1$.  }
%\label{KS3}
\end{figure}  

\begin{figure}[h] 
\includegraphics[width=6.0 in]{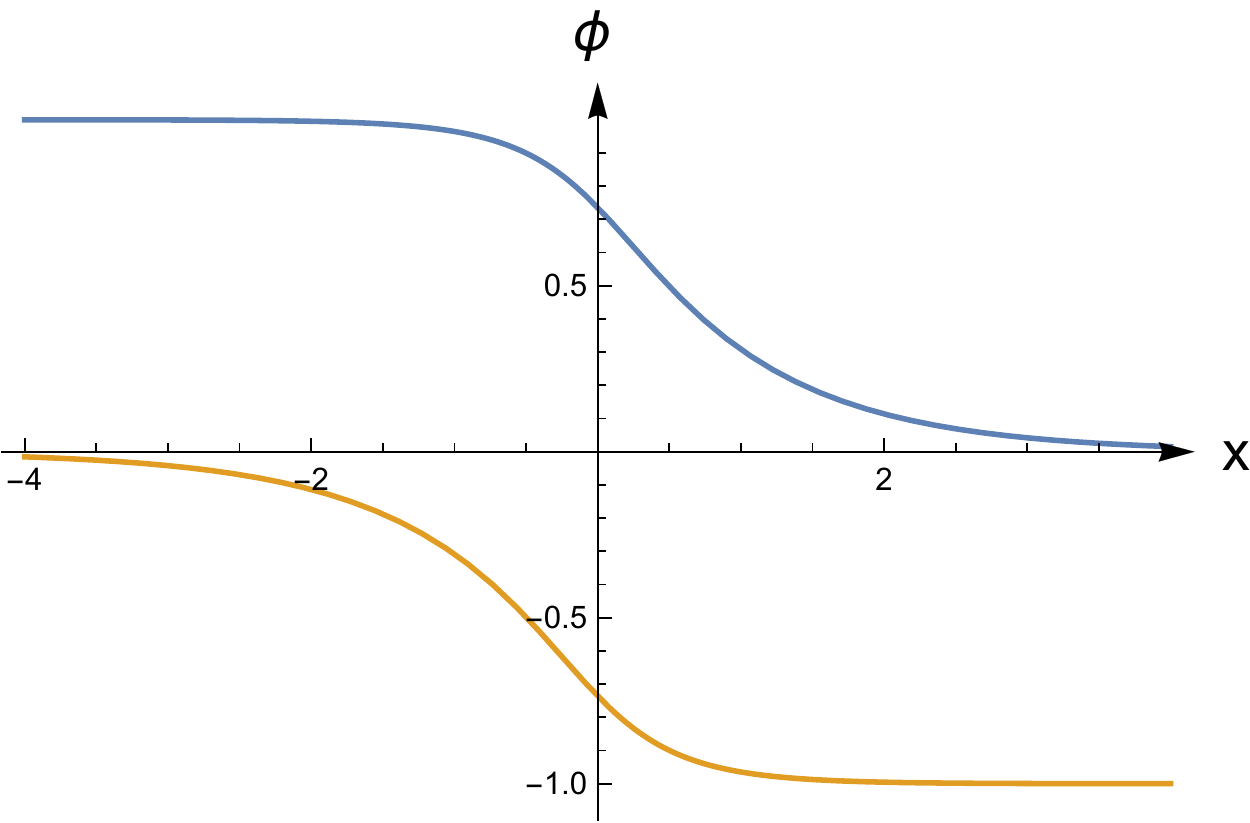}
\caption{Explicit kink solution given by Eq. (38) for $n=1$.  }
%\label{KS1}
\end{figure} 

Unfortunately, at present we do not know how to obtain the kink solution from $1$ 
to $(2)^{1/2n}$, the corresponding mirror kink and the corresponding two antikink
solutions of potential (\ref{4.7}) from the kink solution of the deformed-dual 
potential (\ref{4.6}) since it does not have a kink solution from $1$ to 
$(2)^{1/2n}$.

\vskip 0.1truein  
\noindent{\bf Example II:}

We have recently obtained \cite{ks21} implicit
six kink and six antikink solutions of the model characterized by the potential
\be\label{4.10}
V_1(\phi) = \frac{1}{2} \phi^2 (1-\phi^{2n})^2 (2-\phi^{2n})^2 
(3-\phi^{2n})^2 \,,~~n = 1, 2, 3,... \,,
\ee
without realizing that this potential is deformation-dual to the potential
\be\label{4.11}
V_0(\phi) = \frac{1}{2} \phi^2 (1-\phi^{4n})^2 (2+\phi^{2n})^2\,,
~~n = 1, 2, 3,... \,,
\ee
i.e. the above $V_1(\phi)$ can be obtained from the $V_0(\phi)$ as given by 
Eq. (\ref{4.11}) by applying the deformation function (\ref{1}) on it. Potentials 
given by Eqs. (39) and (40) are shown in Fig. 7 for $n=1$ with seven and 
three degenerate minima, respectively. 
\begin{figure}[h] 
\includegraphics[width=6.0 in]{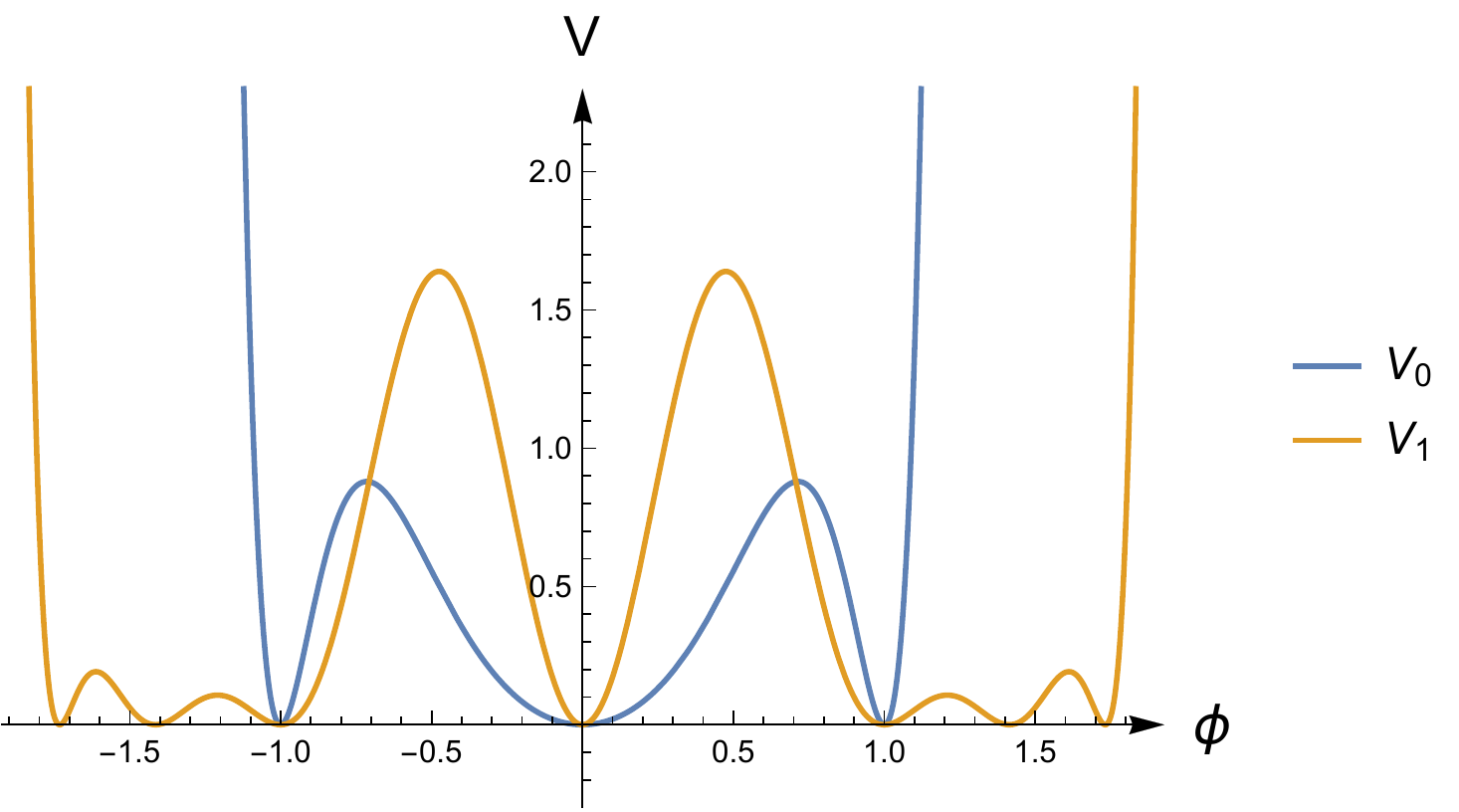}
\caption{Potential $V_1(\phi)$ in Eq. (39) and its dual potential $V_0(\phi)$ in Eq. (40) for $n=1$.  }
%\label{KS3}
\end{figure} 

In order to obtain the kink solution of $V_0(\phi)$ as given by
Eq. (\ref{4.11}) from $0$ to $1$, we need to solve the self-dual equation
\be\label{4.12}
\frac{d\phi}{dx} = \phi (1-\phi^{4n}) (2+\phi^{2n})\,.
\ee
This equation is easily integrated using partial fractions and we obtain
\be\label{4.13}
x^{(0)} = \frac{1}{12} \ln \bigg [\frac{\phi^{6n}(2+\phi^{2n})}
{(1-\phi^{2n}) (1+\phi^{2n})^3} \bigg ]\,.
\ee
It then immediately follows that for the potential $V_0(\phi)$ as given by 
Eq. (\ref{4.11}), the asymptotic behavior of the kink solution from $0$ to
$1$ is given by 
\be\label{4.14}
\lim_{x \rightarrow -\infty} \phi^{(0)}_{K}(x) = (2)^{-1/6} e^{2x/n}\,,~~~
\lim_{x \rightarrow \infty} \phi^{(0)}_{K}(x) = 1- \frac{3}{16n}e^{-12 x}\,.
\ee
The kink solution given by Eq. (42) for $n=1$ is depicted in Fig. 8 with 
both exponential tails clearly visible at the two ends. 
\begin{figure}[h] 
\includegraphics[width=6.0 in]{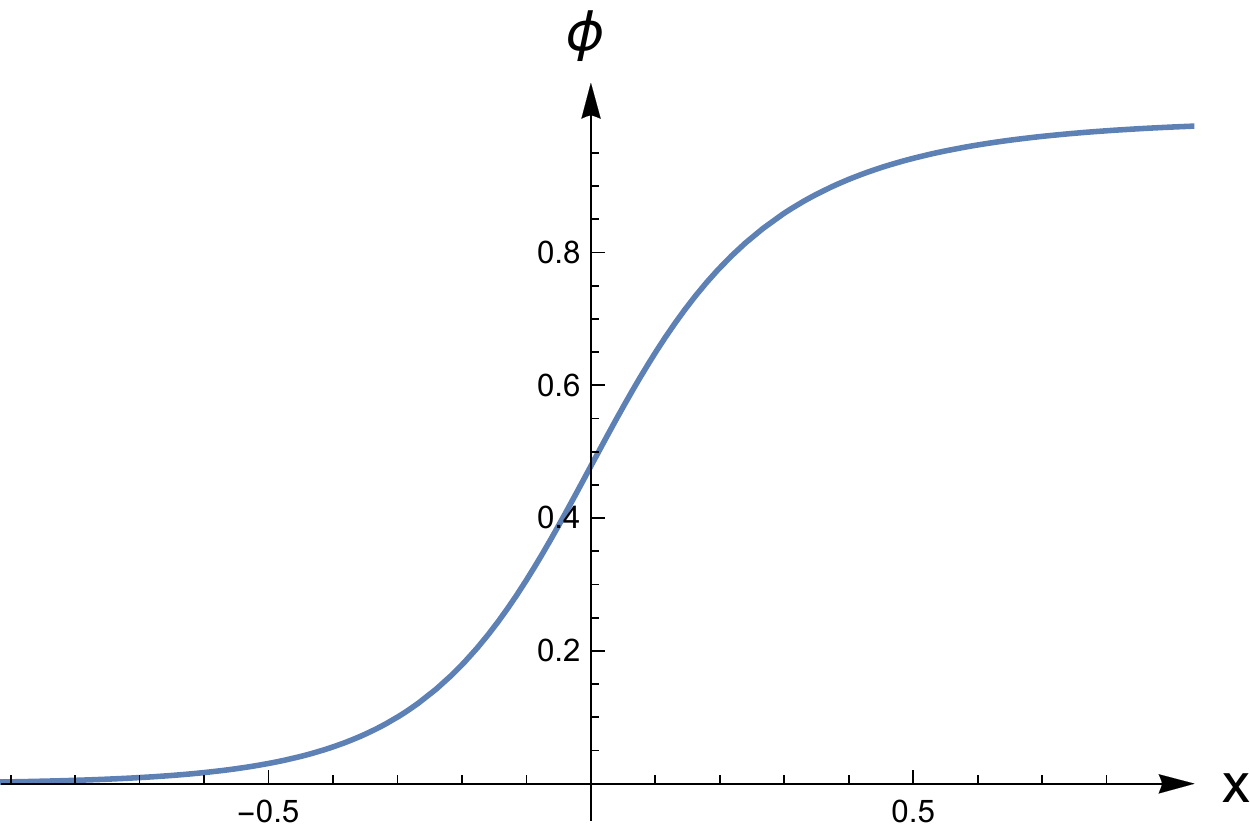}
\caption{Kink solution obtained by inverting implicit Eq. (45) for $n=1$. }
%\label{KS2}
\end{figure}  

\begin{figure}[h] 
\includegraphics[width=6.0 in]{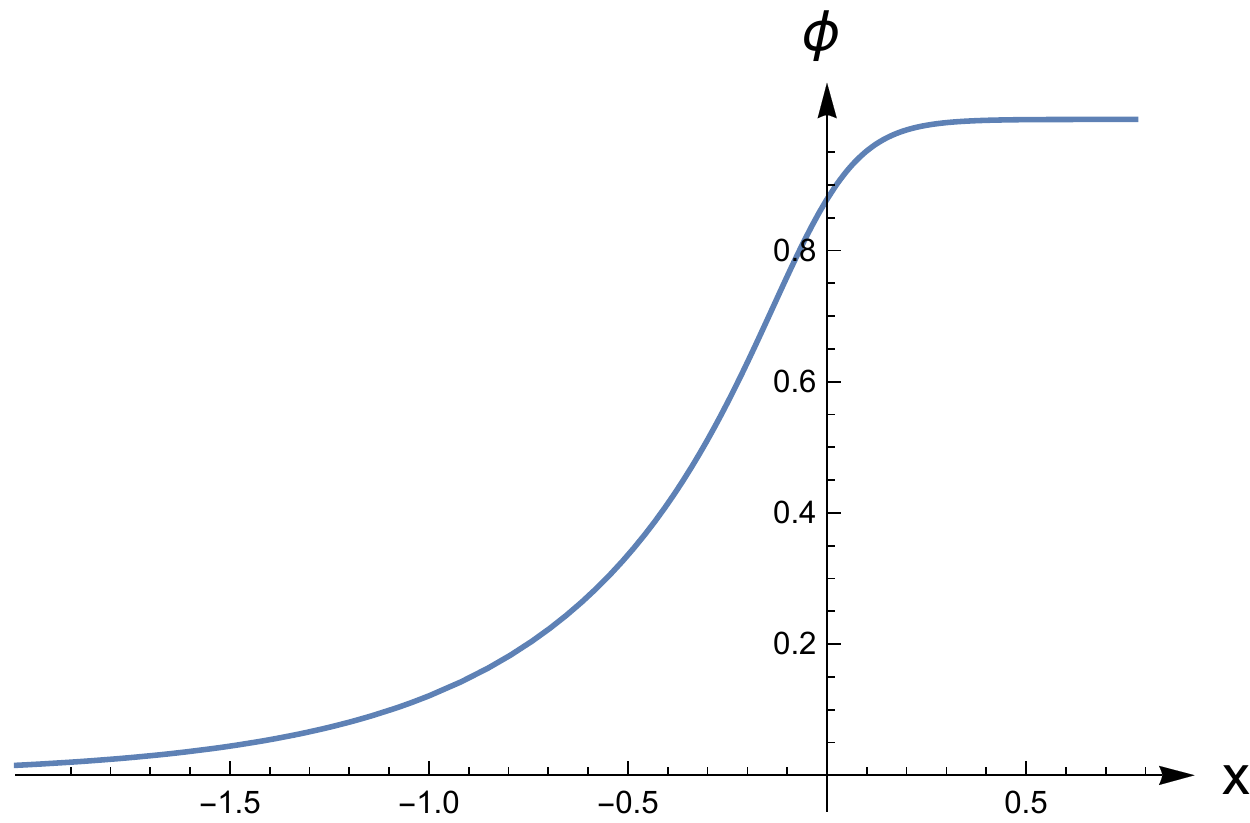}
\caption{Kink solution obtained by inverting implicit Eq. (42) for $n=1$.   }
%\label{KS3}
\end{figure} 

In order to obtain the kink solution of $V_1(\phi)$ as given by
Eq. (\ref{4.10}) from $0$ to $1$, we need to solve the self-dual equation
\be\label{4.15}
\frac{d\phi}{dx} = \phi (1-\phi^{2n}) (2-\phi^{2n}) (3-\phi^{2n})\,.
\ee
This equation is easily integrated using partial fractions and as we have
shown in our recent paper \cite{ks21} we obtain
\be\label{4.16}
x^{(1)} = \frac{1}{12} \ln \bigg [\frac{\phi^{2n} (2-\phi^{2n})^3}
{(1-\phi^{2n})^3 (3-\phi^{2n})} \bigg ]\,.
\ee
It then immediately follows that for the potential $V_1(\phi)$ as given by 
Eq. (\ref{4.10}), the asymptotic behavior of the kink solution from $0$ to
$1$ is given by 
\be\label{4.17}
\lim_{x \rightarrow -\infty} \phi^{(0)}_{K}(x) = (3/8)^{1/2n} e^{6x/n}\,,~~~
\lim_{x \rightarrow \infty} \phi^{(0)}_{K}(x) = 1- \frac{1}{2^{4/3}n}e^{-4 x}\,.
\ee
The kink solution given by Eq. (45) for $n=1$ is depicted in Fig. 9 with 
both exponential tails clearly visible at the two ends. 

Unfortunately, at present we do not know how to obtain the kink solutions from $1$ 
to $(2)^{1/2n}$ and from $(2)^{1/2n}$ to $(3)^{1/2n}$, the corresponding mirror kinks 
and the corresponding four antikink solutions of the potential (\ref{4.10}) from the kink 
solutions of the deformed-dual potential (\ref{4.11}), since it only has a kink solution  
from $0$ to $1$ and the corresponding mirror kink solution.

\section{Conclusions and Open Questions}

In this paper we have discussed several novel properties of the deformation
function as given by Eq. (\ref{1}). We have shown that this deformation
function is its own inverse. As a result, as shown by us, apart from a 
wide class of self-deformed potentials (which are invariant under this
deformation), all other potentials break up into deformed pairs. As also shown
by us, for a wide class of kink bearing potentials, the corresponding deformed-dual 
potential is not bounded from below. On the other hand, there is a wide
class of deformed pair of potentials both of which admit two kink and two
antikink solutions and for which at least one of the kink tails has a power law
tail. In such cases, as shown by us, the appropriate kink tails of the two
deformed pairs with power law tail have different asymptotic behavior. As a
result the kink-kink and the kink-antikink forces in the two deformed pair of
potentials are very different. Finally, we have pointed out what we consider
to be the most remarkable feature of the deformation function (\ref{1}), i.e.
it can act as an arbitrary even number $2m$ of kink creating or kink annihilating
function. That is, starting from a wide class of potentials with two kinks, this deformation 
can take one to the corresponding deformed-dual potential having arbitrary even 
$2m+2$ number of kink solutions. Conversely, by starting from the deformed-dual
potential with $2m+2$ kink solutions and applying the deformed function 
(\ref{1}), one goes to its partner potential having only two kink solutions.

This paper raises several interesting questions, some of which are as follows.

\begin{enumerate}

\item In this paper we have considered a deformation function which is its
own inverse, i.e. naively it is a deformation with period two. 
Generalizing, are there deformation functions of period $2^{n}$, where $n$ is
any positive integer? If yes, what are its key properties? Can it also 
act as a kink creating and/or kink annihilating deformation? How does this
deformation function behave for large $n$?

\item Are there deformation functions of prime number (like $3, 5, 7, 11, ...$) 
period as well as an arbitrary multiple of them? If yes, what are the properties 
of such deformation functions?

\item Are there deformations which can act as a kink creating function with
creating, say $n$, $2n$, $3n$ ... kink solutions or conversely can it act as
a kink annihilating function with reducing $n$, $2n$, $3n$, ... number of kink
solutions?

\item So far the various deformation functions suggested in the literature
are such that the qualitative nature of the kink tail in a given potential
and the corresponding appropriate kink tail of the potential obtained after
the deformation are similar, i.e. both could be exponential or both could
be power law tail or other tails, such as super-exponential \cite{pks}, etc. 
Is there a deformation for which the qualitative nature of the appropriate 
kink tails are qualitatively different? 

\end{enumerate}

It would be quite insightful if one could answer some of the questions raised here.

\vskip 0.1 in

\noindent{\bf Acknowledgment:} We thank Ayhan Duzgun for the help with the 
figures. A.K. is grateful to Indian National Science Academy for the award of INSA 
Senior Scientist position at Physics Department,Savitribai Phule Pune University, India. 
The work at Los Alamos National Laboratory (A.S.) was carried out under the auspices 
of the U.S. DOE and NNSA under Contract No. DEAC52-06NA25396.

\end{document}